\documentstyle[preprint,aps]{revtex}

\tightenlines

\begin{document}

\draft

\title{\bf Density Matrix Renormalization Group Approach to 
the Massive 
Schwinger Model}
\author{T. Byrnes, P. Sriganesh, R.J. Bursill, and C.J. Hamer.
}
\address{School of Physics, 
The University of New South Wales, 
Sydney, 
NSW 2052, 
Australia.}                      

\date{\today}

\maketitle 

\begin{abstract}
The massive Schwinger model is studied, using a
density matrix renormalization group approach to the staggered lattice
Hamiltonian version of the model. Lattice sizes up to 256 sites are
calculated, and the estimates in the continuum limit are almost two
orders of magnitude more accurate than previous calculations. 
Coleman's picture of `half-asymptotic' particles at background 
field $\theta = \pi$ is confirmed. The predicted
phase transition at finite fermion mass $ (m/g) $ is accurately 
located, and
demonstrated to belong in the 2D Ising universality class.
\end{abstract}                    

\narrowtext

The Schwinger model\cite{schwinger62}, or
quantum electrodynamics in one space and one time dimension, exhibits
many analogies with QCD, including
confinement, chiral symmetry
breaking, charge shielding, and a topological
$\theta$-vacuum\cite{casher74,coleman75,coleman76}. It is a common 
test bed for the trial of new techniques for the study of QCD:
for instance, several authors
 have discussed new methods
of treating lattice fermions using the Schwinger model as an example
\cite{creutz95}.

	Our purpose in this paper is twofold. First, we aim to explore
the physics of this model when an external `background' electric field
is applied, as discussed long ago in a beautiful paper by
Coleman\cite{coleman76}. Secondly, we wish to demonstrate the
application of density matrix renormalization group 
(DMRG) methods \cite{white92,gehring97} to a
model of this sort, with long-range, non-local Coulomb interactions.
The DMRG approach has been used with great success for lattice spin
models and lattice electron models such as the Hubbard model; here we
apply it to a lattice gauge model.

Choosing a time-like axial gauge
$A_0 = 0$,
the Schwinger model Hamiltonian becomes
\begin{equation} \label{fermionham}
 H = \int dx \left( - i \bar{\psi} \gamma^1 (\partial_1 + i g A_1 ) 
\psi + m \bar{\psi} \psi + \frac{1}{2} E^2 \right).
\end{equation}
where $ \psi $ is a 2-component spinor field, since there is no 
spin in one 
space dimension. The coupling $ g $ has dimensions of mass, so 
the theory is 
super-renormalizable. Using $ g $ as the scale of energy, the 
physical 
properties of the model are then functions of the 
dimensionless ratio $ m/g $. 
Gauss' law becomes
\begin{equation}
\partial_1 E = - \partial_1 \dot{A}^1 = g j^0 = g 
\bar{\psi} \gamma^0 \psi.
\label{guass}
\end{equation}
where $E$ is the 1-component electric field.
This can be integrated to give
\begin{equation}
E = g \int dx j^0(x) + F,
\end{equation}
showing that $ E $ is not an independent field, but can be 
determined in terms 
of the charge density $ j^0(x) $, up to the constant of 
integration $ F $, 
which corresponds to a ``background field'', as discussed by 
Coleman \cite{coleman76}. We can think of the background field 
as created by condenser plates at either end of our one-dimensional 
universe.

If $ |F| > g/2 $, charged pairs will be produced, and separate 
to infinity, until the field is reduced within the range 
$ |F| \leq g/2 $, thus lowering the electrostatic energy per 
unit length. Thus physics is periodic in $ F $  with period 
$ g $, and it is convenient to define an angle $ \theta $ by
$\theta = 2 \pi F/g$.
Then we can always choose $ \theta $ to lie in the interval 
$ [0,2 \pi ] $.

In the weak-coupling limit $ m/g \rightarrow \infty $, the vacuum 
contains no 
fermionic excitations, and the vacuum energy density 
$ \epsilon_0 $ is given 
purely by the electrostatic energy term (we ignore the energy of 
the Dirac sea)
$\epsilon_0 =  
F^2/2  $ $ (\theta \leq \pi)$
or $ (g-F)^2/2 $ $ (\pi \leq \theta \leq 2 \pi)$.

Thus there is a discontinuity in the slope of the energy density, 
corresponding
 to a first-order phase transition, at $ \theta = \pi $. In the 
strong-coupling
 limit $ m/g = 0 $, on the other hand, chiral invariance demands 
that the 
vacuum energy density remains {\it constant} as a function of 
$ \theta 
$\cite{coleman76}. Thus 
we expect a first-order transition at $ \theta = \pi $ for large 
$ m/g $, which
 terminates at a second order critical point at some finite 
$ (m/g)_c $.
The two-fold vacuum degeneracy would naturally lead one to expect 
that the transition will lie in the Ising universality class.

\begin{figure}
\centerline{\psfig{file=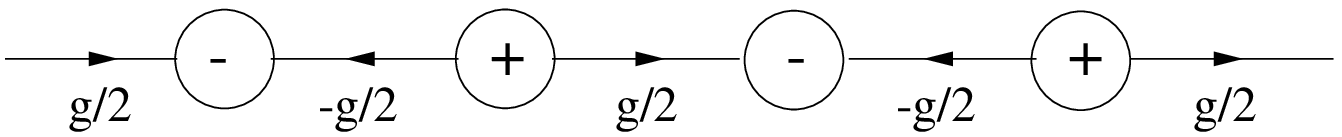,clip=}}
\caption{A configuration of ``half-asymptotic'' charged fermions 
at background field $ F = g/2 $.}
\label{fig:halfass}
\end{figure}

Normally, charge is confined in the model: there is a `string' of 
constant 
electric field (or flux) connecting any pair of opposite charges 
\cite{casher74,coleman75}. But Coleman \cite{coleman76} points out 
that in the 
very special case $ \theta = \pi $, the peculiar phenomenon 
of ``half-asymptotic'' particles arises. In the weak-coupling limit, 
one can 
envisage the state shown in Fig. \ref{fig:halfass}. The electric 
field energy 
density is the same in between each pair of particles, and they can 
therefore 
move freely, as long as they maintain the same {\it ordering}, i.e. 
no pair of 
fermions interchanges positions. 

The one-dimensional fermionic theory can be mapped into an 
equivalent Bose 
form \cite{coleman75,mandelstam75}. 
At $m/g = 0$, the theory is exactly solvable, and reduces 
to a theory of
free, massive bosons, with mass $M = g/\sqrt{\pi}$, independent 
of the
background field $\theta$. At large $m/g$ and $\theta = \pi$, 
where the
spontaneous symmetry breaking occurs, the half-asymptotic particles 
are
found to correspond to topological solitons, or ``kinks" and
``antikinks", in the Bose language \cite{coleman76}.

	There have been very few attempts to verify these predictions 
numerically, that we are aware
of. Hamer, Kogut, Crewther and Mazzolini\cite{hamer82} 
used finite-lattice techniques to address the problem, and
located the phase transition at $\theta = \pi$ to lie at 
$(m/g)_{c} =
0.325(20)$, with a correlation length index $\nu = 0.9(1)$. 
Schiller and Ranft \cite{schiller82} used Monte Carlo techniques to
locate the phase transition at $(m/g)_{c} = 0.31(1)$.
	 For zero background field, on the other hand, many different 
numerical methods have been applied to the
Schwinger model;
for a recent review, see Sriganesh {\it et al.}
\cite{sriganesh00}.

We employ the Kogut-Susskind \cite{kogut75} Hamiltonian lattice 
formulation of 
the Schwinger model, following Banks {\it et al.} \cite{banks76}. 
The
dimensionless lattice Hamiltonian is
\begin{equation}
W = \frac{2}{a g^2} H  =
  \sum_n L^2 (n) + \mu \sum_n (-1)^n \phi^\dagger (n) \phi (n)
-ix \sum_n \left[ \phi^\dagger (n) e^{i \theta(n) } \phi(n+1)
- \mbox{h.c.} \right]
\label{latticeham}
\end{equation}
where $a$ is the lattice spacing, $\phi(n)$ is a single-component
fermion field at site $n$, $e^{i\theta(n)}$ is the link variable 
at link
 $ n $, and
$\mu = 2m/(g^2 a)$, $x = 1/(g^2 a^2)$.

We use a ``compact'' formulation where the gauge field becomes an 
angular 
variable $ 0 \leq \theta(n) \leq 2 \pi $ on the lattice, and 
$ L(n) $ is the 
conjugate spin variable (the lattice equivalent of the electric field)
\begin{equation}
[ \theta(n), L(m) ] = i \delta_{nm}
\end{equation}
so that $ L(n) $ has integer eigenvalues 
$ L(n) = 0, \pm 1, \pm 2, \dots $. 

In the lattice strong-coupling limit $ x \rightarrow 0 $, the 
unperturbed ground state $ | 0 \rangle $ has
\begin{equation}
\label{groundalpha0}
L(n) = 0, \ \ \ \ \ 
\phi^\dagger(n) \phi(n) = \frac{1}{2} [ 1 - (-1)^n ], \ \ \ \ \ 
\mbox{all } n
\end{equation}
whose energy we normalize to zero. The lattice version of Gauss' 
law is then taken as
\begin{equation} \label{latticegauss}
L(n) - L(n-1) = \phi^\dagger (n) \phi(n) - \frac{1}{2}[1-(-1)^n]
\end{equation}
which means excitations on odd and even sites create $ \mp 1 $ 
units of flux, corresponding to ``electron'' and ``positron'' 
excitations respectively. Eq. (\ref{latticegauss}) determines 
the electric field $ L(n) $ entirely, up to an arbitrary additive 
{\it constant} $ \alpha $, which then represents the background 
field. Allowing $ \alpha $ to be non-zero, the electrostatic energy 
term is modified to
$\sum_n L^2 (n) \rightarrow \sum_n (L(n) + \alpha)^2 $.
The physics of the background field then matches precisely with the 
continuum 
discussion. 
Physics is then periodic in $ \alpha $ with period 1, and the 
background field variable is
$\theta = 2 \pi \alpha$ .

In the weak-coupling limit $ m/g \rightarrow \infty $ and for 
background
field $\alpha = 1/2$, there are two degenerate vacuum or `loop' 
states,
$|\frac{1}{2} \rangle$ and $|-\frac{1}{2} \rangle$, corresponding 
to $\{L(n) + \alpha = \pm 1/2, \mbox{ all } n \}$,
respectively. On a finite lattice, the eigenstates will be the 
symmetric
and anti-symmetric combinations of those, and we will denote the 
energy
gap between them by $\Delta_{0}$. The single-fermion state will 
consist
of a single electron with $L(n) + \alpha = +1/2$ to its left and
$-1/2$ to its right; the energy gap between this and the ground state
will be denoted $\Delta_{1}$. Finally, the energy gap to the lowest
2-particle electron-positron state will be denoted $\Delta_{2}$.

We will also study two order parameters which can be used to
characterize the phase transition at $\theta = \pi$. The first 
one is
the average electric field
\begin{equation}
\Gamma^{\alpha} = \frac{1}{N} \langle \sum_{n}(L(n) + \alpha) 
\rangle_{0},
\end{equation}
which in the weak-coupling limit $m/g \rightarrow \infty$ takes 
values
$\pm 1/2$ for the zero-loop and one-loop states respectively. 
	The second order parameter is the axial fermion density
suggested by Creutz\cite{creutz95}
\begin{eqnarray}
\Gamma^{5} & = & \langle i \bar\psi\gamma^{5}\psi/g \rangle_{0} \\ 
 &  = &  - \frac{i \sqrt{x}}{N} \langle 
\sum_{n} (-1)^{n} \left[ \phi^{\dagger}(n)\phi(n+1)
- \mbox{h.c.} \right] \rangle_{0}.
\end{eqnarray}

Our results are based on the density matrix renormalization group 
(DMRG) 
method \cite{white92,gehring97}. 
The method employed here is the ``infinite system'' DMRG method, 
as prescribed 
by White \cite{white92}, used both with open and periodic boundary 
conditions
(OBC and PBC).  Due to the presence of the electric field on the 
links, and the
 differing nature of the odd and even-numbered sites, some 
modifications were 
made to the method, although nothing that changes the spirit of the 
DMRG. 

For a particular spin configuration with OBC, 
if we specify the incoming electric field for the first site of the 
chain, 
then according to Gauss's law the electric field for all the links 
can 
be deduced. 
If PBC are imposed, then as there is no particular link 
to fix the electric field, we can have loops of electric flux 
extending 
throughout the ring. In the presence of a background field we 
simply add (or 
subtract) $\alpha$ from the values of the electric fields due to 
the spin 
configuration. A cutoff, or maximum loop value was chosen such that 
full convergence was 
reached to machine precision. A loop range of [-5,5] was more than 
sufficient 
in most cases.

In the standard ``infinite system'' method one splits the chain 
(or ring) 
into two blocks and two sites, and in each DMRG iteration the blocks 
increase 
in size by a single site. Due to the differing nature of the 
odd and even sites in the lattice formulation (\ref{latticeham}), we 
modify 
this so that two sites are augmented each time, so that the 
superblock grows 
by four sites in a single DMRG iteration. 
We truncate the basis in each DMRG iteration such that there are a 
maximum of 
125 states per spin sector per loop in a block. 

The accuracy of the DMRG calculation is strongly dependent on the 
parameter 
values $ x $ and $ m/g $. 
In the worst case scenario 
$ (m/g = 0.3; x = 100; \theta = \pi; N = 256) $ 
with PBC, 
which represents the smallest lattice spacing used for this study, 
and lies in 
the critical region,  
 the ground state energy can be resolved 
to 1 part in $10^{6}$, while the two-particle gap is resolved to 1 
part
in $10^{3}$, while the order parameters are accurate
to 3 figures. 
It is more usual to obtain ground state eigenvalues to near machine 
precision, 
and ``vector'' gap errors of 1 part in $ 10^6 $.  

\begin{figure}
\centerline{\psfig{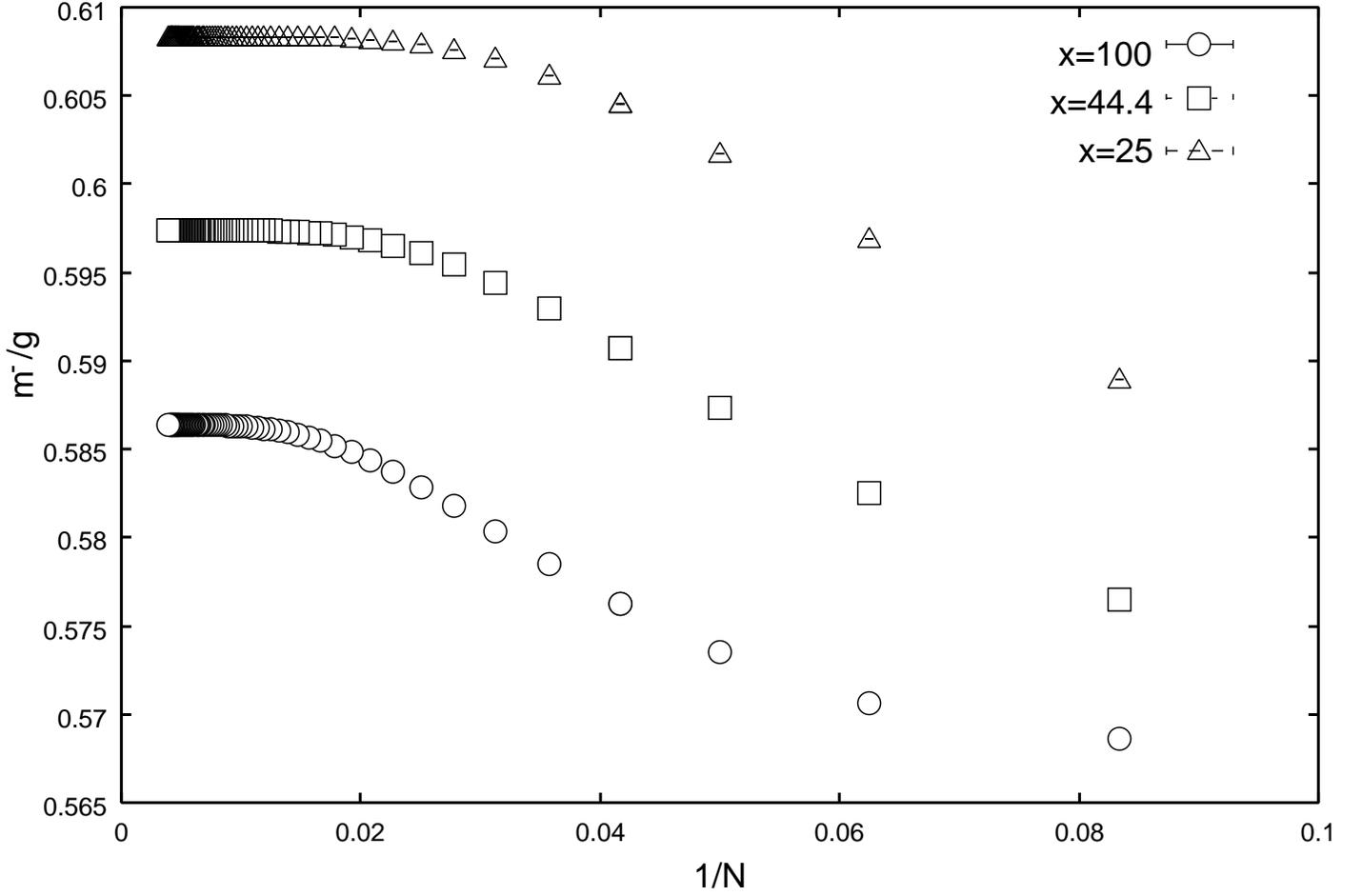}}
\caption{``Vector'' mass gaps for $ m/g = 0 $, finite lattices 
$ N = 10$-$ 256 $, and various lattice spacings.}
\label{fig:vectorm0}
\end{figure}

We begin with the case of zero background field, $\theta = 0$, in order
to demonstrate the power of the DMRG method.
The most accurate results to date are those of Sriganesh {\it et al.}
\cite{sriganesh00}, who used exact diagonalization on lattices up 
to $N = 22$ sites.
With DMRG it is possible to go to $N = 256$ sites, such that there
is essentially no extrapolation necessary to obtain the bulk limit $ N
\rightarrow \infty$. 
Fig. \ref{fig:vectorm0} shows the data for $ m/g = 0 $, which is 
the exactly 
soluble case. 
For $ x = 4 $ through to $ x = 100 $ we 
have 6 figure convergence. 
The final continuum limit $a \rightarrow 0$ or $x \rightarrow
\infty$  
is obtained by making polynomial fits in powers of $1/\sqrt{x} = ga$ 
as 
used in Ref. \cite{sriganesh00}. 
Our final estimate of $ m^-/g = 0.56419(4) $
for this case agrees extremely well with the analytic result 
of $1/\sqrt{\pi} = 0.5641896..$, and is about 25 times more accurate
than the corresponding estimate $0.563(1)$ by 
Sriganesh {\it et al.} \cite{sriganesh00}.

\begin{figure}
\centerline{\psfig{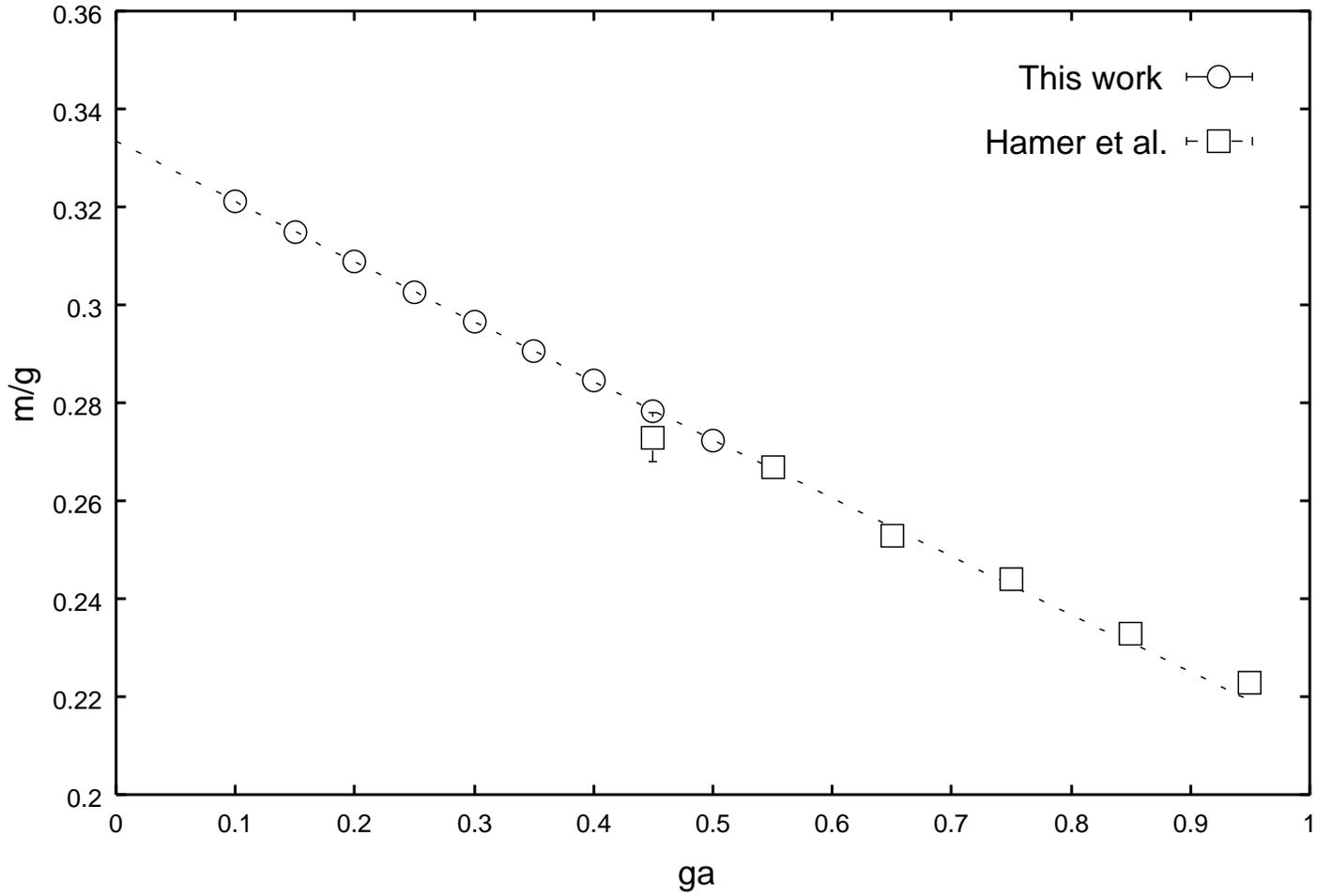}}
\caption{Critical line in the $ m/g $ versus $ 1/\sqrt{x} = ga $ 
plane. Open circles are our present estimates, and squares are 
the previous results of Hamer {\it et al.} \protect\cite{hamer82}, 
which are in good agreement. The dashed line is a quadratic fit to 
the data in $ ga $. }
\label{fig:finalplot}
\end{figure}

We now turn to the case of background field $\theta = \pi$. First, 
we use 
finite-size scaling theory to estimate the position of the 
critical point, by 
calculating ``pseudo-critical points" at each lattice size $ N $, 
and each lattice spacing $ x $. 
The pseudo-critical point is defined\cite{barber83} as the 
point where two successive
finite-lattice energy gaps scale like $1/N$.
For this exercise we used the `loop
gap' $\Delta_{0}/g$, which collapses to zero at the
 pseudo-critical point. 
The pseudo-critical points converge very rapidly to the bulk 
limit, like
$1/N^{3}$.
The results are  
 shown in Fig. \ref{fig:finalplot}. 
 A quadratic fit in $ 1/\sqrt{x} $ extracts 
the continuum limit, which we estimate to be
$(m/g)_{c} = 0.3335(2)$.
This is consistent with the previous estimates by 
Hamer {\it et al.} \cite{hamer82} of $ (m/g)_c
 = 0.325(20) $, or Schiller and Ranft \cite{schiller82}, 
$(m/g)_{c} =
0.31(1)$, but with two orders of magnitude improvement in accuracy. 

\begin{figure}
\centerline{\psfig{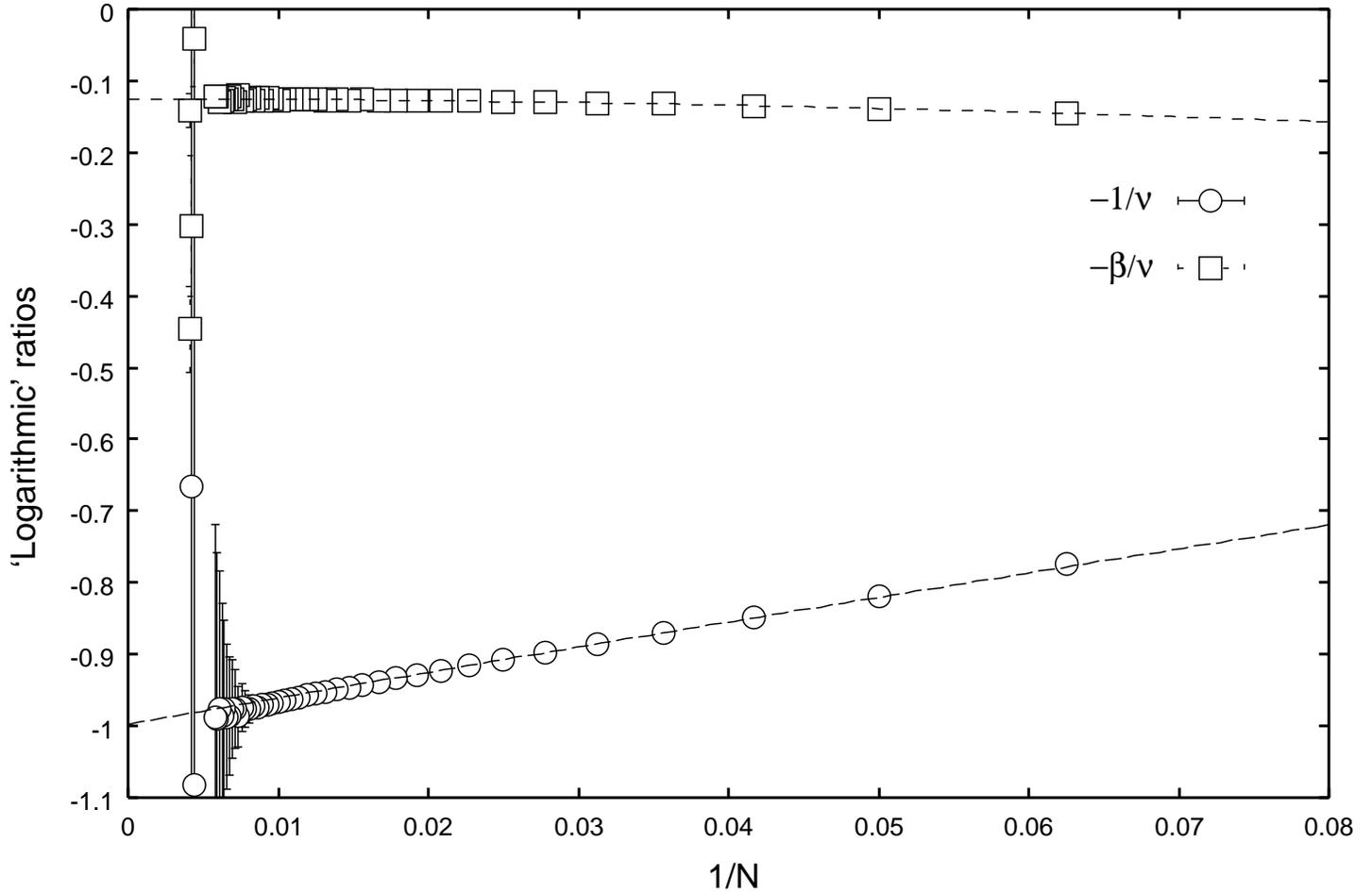}}
\caption{`Logarithmic' ratio estimates of critical indices 
$ - 1/\nu $ and $ - \beta / \nu $ for lattice spacing 
$ 1/\sqrt{x} = ga = 0.45 $. Quadratic fits in $ 1/N $ provide 
the bulk extrapolations. We estimate here $ 1/\nu = 1.00(2) $ 
and $ \beta/\nu = 0.125(5) $.}
\label{fig:ratios}
\end{figure}

Finite-size scaling theory  \cite{barber83} also allows us to 
estimate the 
critical indices for the model. 
The order parameters $ \Gamma^\alpha_N $ and $ \Gamma^5_N $, 
for instance, are expected 
to scale like 
$ \sim N^{-\beta/\nu} $ \cite{barber83}.  
Then the logarithmic ratios 
\begin{equation}
\frac{ \ln [ \Gamma_N ( (m/g)_N^* ) / \Gamma_{N-1} ( (m/g)_N^* ) ] }
{ \ln [ N/(N-1) ] } \sim - \frac{\beta}{\nu} 
\end{equation}
give a direct estimate of the index ratio.
 Fig. \ref{fig:ratios} shows the resulting estimates for 
$ \beta/\nu $, 
obtained from the electric field order parameter, and for 
$1/\nu$, obtained from
the beta function.
 We see essentially no variation in the exponents with 
lattice spacing, to within
the accuracy of our calculations. 
Our best estimates for the 
critical exponents are thus
$\nu  =  0.99(1)$, $ \beta/\nu  =  0.125(5) $.
These results provide reasonably conclusive evidence that the 
Schwinger model
transition at 
$ \theta = \pi$ lies in the same universality class as the one 
dimensional 
transverse Ising model, or equivalently the 2D Ising model, 
with ($ \nu = 1, 
\beta = 1/8 $).

\begin{figure}
\centerline{\psfig{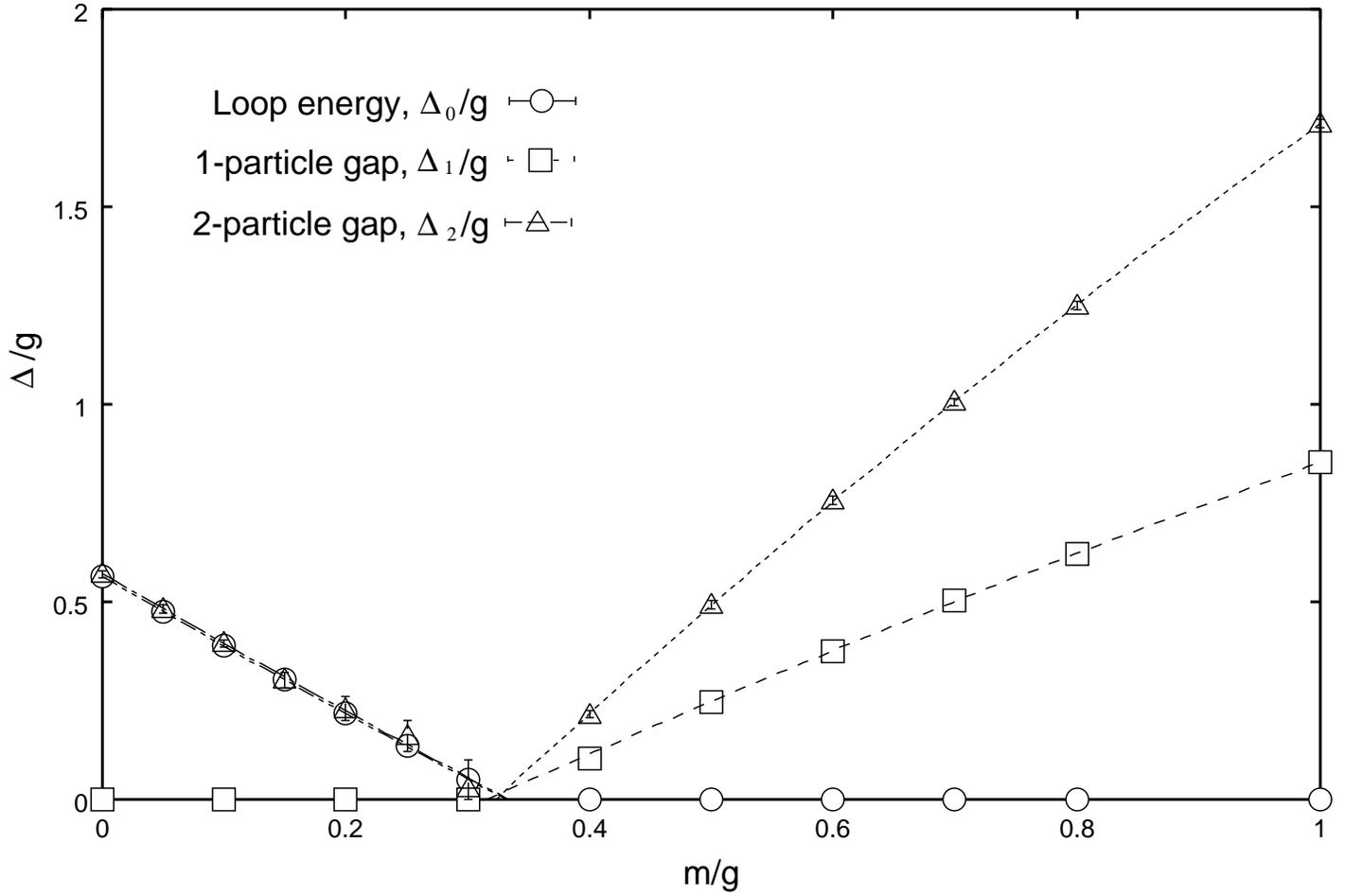}}
\caption{Final estimates for gaps in the 0-particle, 1-particle and
2-particle sectors at $ \theta = \pi $. Dashed lines are merely to 
guide the eye.}
\label{fig:mogland}
\end{figure}

We now turn to estimating continuum limit values for the energy 
gaps and 
order parameters.  
The convergence at $\theta = \pi$ is not so good as at $\theta = 0$.
Fig. \ref{fig:mogland} displays our final results for the energy gaps
$\Delta_{0}/g$ between the ``loop" states, $\Delta_{1}/g$ 
in the 1-particle
sector, and $\Delta_{2}/g$ in the 2-particle sector, for all 
values of $ m/g $.
 
We see that all gaps vanish at the critical point. 
The `loop' gap is zero for all $ m/g > 
(m/g)_c $, as predicted by Coleman \cite{coleman76}. 
The 2-particle gap $\Delta_{2}/g$ 
vanishes at the critical point $ (m/g)_c $, but on either 
side of the critical point 
there is a finite gap, and an
almost linear behavior with $ m/g $. 
Note,
however, 
that the behavior is not {\it exactly} linear. 
The 1-particle gap 
vanishes for $ m/g < (m/g)_c $,
while 
for $ m/g > (m/g)_c $ it is very close to half the 2-particle 
gap. 
Once again, the behavior is very nearly linear in $m/g$.

	The pattern of eigenvalues exhibited in Fig. 
\ref{fig:mogland} bears
an extraordinary resemblance to that of the transverse Ising
model \cite{fradkin}.
In particular, we see that the energy of the 1-particle 
or `kink'
state vanishes at the critical point, and then 
{\it remains degenerate}
with the ground state for $(m/g) < (m/g)_{c}$. Assuming this degeneracy
is exact, this indicates that a `kink condensate' will form in the
ground state for small mass, as discussed by Fradkin and
Susskind\cite{fradkin}. It also indicates the existence of a new `dual
symmetry' in the model, 
which is spontaneously broken in the low-mass region, and has not been 
explored hitherto. There should also be a `dual order
parameter' associated with this symmetry.

\begin{figure}
\centerline{\psfig{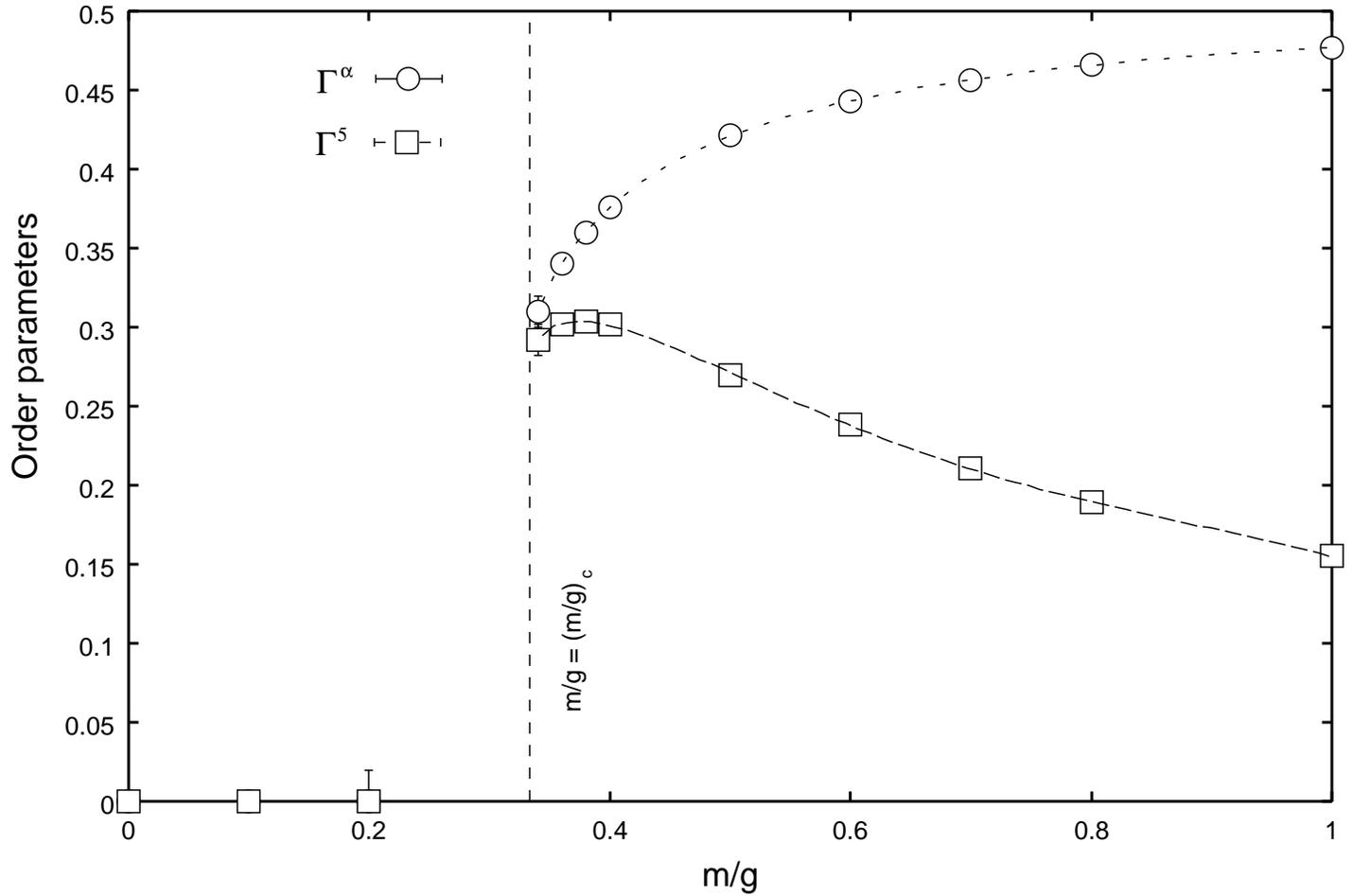}}
\caption{Order parameters  
$ \Gamma^{\alpha} = \langle (L + \alpha) \rangle_0 $
, $ \Gamma^{5} = \langle i \bar{\psi} \gamma_5 \psi /g \rangle_0  $ 
near the critical region. Dashed lines are merely to guide the eye.}
\label{fig:moglandorder}
\end{figure}

We may also obtain estimates for the order parameters $\Gamma^5$ and
$\Gamma^\alpha$ as functions of $m/g$. Our results are 
displayed in Fig. 
\ref{fig:moglandorder}. Both order parameters are zero, 
within errors, for 
$ m/g < (m/g)_c $. 
It appears that both order parameters turn over and drop abruptly to
zero as the critical point is approached from above, 
consistent with the
small exponent $ \nu = 1/8 $ found previously.
The axial density $\Gamma^5$ decreases steadily towards zero at large
$m/g$, whereas $\Gamma^\alpha$ approaches the expected value of $1/2$.
More detailed results will be given in a forthcoming
paper\cite{byrnes02}.


We would like to thank Profs. Michael Creutz and Jaan Oitmaa and Dr.
Zheng Weihong for
useful discussions and help.
We are grateful for computational facilities provided by 
The New South Wales 
Center
for Parallel Computing, The Australian Center for Advanced Computing
and Communications and The Australian Partnership for 
Advanced Computing.
R.B. was supported by The Australian Research Council and The J. G.
Russell Foundation.

\center
\widetext
\input psfig
\psfull


\begin{references}


\bibitem{schwinger62} J. Schwinger, Phys. Rev. {\bf 128}, 2425 (1962).


\bibitem{casher74} A. Casher, J. Kogut, and L. Susskind, 
Phys. Rev. D {\bf 10}, 732 (1974).

\bibitem{coleman75} S. Coleman, R. Jackiw, and L. Susskind,
   Ann. of Phys. {\bf 93}, 267 (1975).


\bibitem{coleman76} S. Coleman, Ann. of Phys. {\bf 101}, 239 (1976).

\bibitem{creutz95} M. Creutz, Nucl. Phys. Proc. Suppl. {\bf 42}, 
56 (1995).


\bibitem{white92} S.R. White, Phys. Rev. Lett. {\bf 69}, 
2863 (1992); Phys. Rev. B {\bf 48}, 10345 (1993).

\bibitem{gehring97}
G.A. Gehring, R.J. Bursill and T. Xiang, Acta Phys. Pol. {\bf 91}, 105 
(1997).

\bibitem{mandelstam75} S. Mandelstam, Phys. Rev. D {\bf 11}, 
3026 (1975).

\bibitem{hamer82} C.J. Hamer, J. Kogut, D.P. Crewther, 
and M.M. Mazzolini, Nucl. Phys. {\bf B208}, 413 (1982).

\bibitem{sriganesh00} P. Sriganesh, C.J. Hamer and R.J. Bursill, 
Phys. Rev. D {\bf 62}, 034508 (2000). 

\bibitem{schiller82}
A.J. Schiller and J. Ranft, Nucl. Phys. {\bf B225}, 204 (1983).

\bibitem{kogut75} J. Kogut and L. Susskind, Phys. Rev. D {\bf 11},
395 (1975).

\bibitem{banks76} T. Banks, L. Susskind, and J. Kogut, 
Phys. Rev. D {\bf 13}, 1043 (1976).

\bibitem{byrnes02} T. Byrnes {\it et al.}, in preparation.

\bibitem{barber83} M.N. Barber, in {\it Phase Transitions 
and Critical Phenomena}, vol. 8, edited by C. Domb and 
J. Lebowitz (Academic, New York, 1983).


\bibitem{fradkin} E. Fradkin and L. Susskind, 
Phys. Rev. {\bf D17}, 2637 (1978).



\end{references}
\end{document}